# Deciphering water-solid reactions during hydrothermal corrosion of SiC

Jianqi Xi[1*], Cheng Liu[1], Dane Morgan[1], Izabela Szlufarska[1,2*]

[1.] Department of Materials Science and Engineering, University of Wisconsin, Madison, WI, 53706, USA.

[2.] Department of Engineering Physics, University of Wisconsin, Madison, WI, 53706, USA

**Abstract**

Water-solid interfacial reactions are critical to understanding corrosion. More specifically, it is notoriously difficult to determine how water and solid interact beyond the initial chemisorption to induce the surface dissolution. Here, we report atomic-scale mechanisms of the elementary steps during SiC hydrothermal corrosion – from the initial surface attack to surface dissolution. We find that hydrogen scission reactions play a vital role in breaking Si-C bonds, regardless of the surface orientations. Stable silica layer does not form on the surface, but the newly identified chemical reactions on SiC are analogous to those observed during the dissolution of silica. SiC is dissolved directly into the water as soluble silicic acid. The rate of hydrothermal corrosion determined based on the calculated reaction activation energies is consistent with available experimental data. Our work sheds new light on understanding and interpreting the experimental observations and it provides foundation for design of materials that are resistant to corrosion in water.

**Keywords:** Atomistic modeling, Corrosion, Rate-limiting step

## 1. Introduction:

Corrosion is one of the most common concerns for materials operating at elevated temperatures or aqueous environments. Corrosion is generally defined as the deterioration of a material or its properties due to the reactions with the surrounding environments [1–5]. Although corrosion has been long recognized as an important degradation process and numerous investigations have brought in many insights into corrosion mechanisms and preventive strategies, control and prediction of how a material will degrade in a given environment remains a formidable challenge [4]. Thermodynamic considerations provide a powerful way to explain corrosion phenomena [2–4], including calculations of redox potential to quantify the driving force of corrosion [2]. However, as discussed in the past decades, corrosion of materials in aqueous solutions is often thermodynamically favorable but not detected in practice [4]. Therefore, in addition to thermodynamic analysis, it is important to understand the kinetics of processes that drive corrosion, which can be done by elucidating elemental steps of interfacial reactions.

Silicon carbide (SiC) is known to exhibit many outstanding properties, such as high-temperature strength, wide band gap, superior radiation tolerance, and oxidation resistance, making it a versatile material for multiple industrial applications. In many of the potential applications of SiC, corrosion is a potential issue that can limit the lifetime of SiC-based engineering components. For instance, in bio-sensor applications, the potential damage of surfaces, due to the presence of bio-objects and the reaction with organic tissue in aqueous environment, could result in a chronic deterioration effect, reducing the longevity of the devices [5–7]. In the nuclear industry, hydrothermal corrosion is a known concern for application of SiC as cladding [8–11], as it has been shown that exposure to high-purity and high-temperature water leads to the corrosion of SiC. For example, Henager Jr *et al*. [11] observed pitting corrosion of SiC when it was exposed to high temperature water (573 K and 10 MPa). By using scanning electron microscope equipped with energy dispersive spectroscopy, the authors observed that carbon-rich pits formed on SiC within five months of the exposure. These results suggest that silicon is easier to dissolve from SiC than carbon during hydrothermal corrosion. In addition, Henager Jr *et al*. [11] suggested that the pitting corrosion in their experiments could be associated with surface impurities, inclusions etc., which could all initiate heterogeneous corrosion. Localized corrosion with the formation of a carbon-rich film under hydrothermal corrosion has also been reported by Terrani *et al*. [8] and Kim *et al*. [9, 10]. However, in this case the carbon-rich pits were found to be unstable in the hydrothermal environment and they were eventually found to dissolve as gas phases after a long-time exposure [11]. These results suggest that the observed pitting corrosion on the SiC surface may be a transient phenomenon and the corrosion could eventually move toward more uniform degradation of the surface. It is interesting to point out that the observed dissolution of carbon-rich region in high temperature water is in contrast to corrosion of SiC in molten salt environments, where a carbon-rich layer forms and remains on the surface as a potentially protective layer for continued corrosion [12]. It remains to be determined to what extent the carbon-rich region provides a passive protection to SiC during hydrothermal environment.

---

[*] *Department of Materials Science and Engineering, University of Wisconsin, Madison, WI, 53706, USA.*
E-mail: szlufarska@wisc.edu, jxi4@wisc.edu; Tel: +1-608-265-5878

Hydrothermal corrosion of SiC is commonly observed in applications, yet it remains unclear as to why and how it occurs, which limits the ability to control it. Specifically, although the carbon species on a corroding SiC surface are expected to be finally dissolved as gas [8–11], the mechanisms of the initial dissolution of silicon species, which have been postulated to play a significant role in the degradation of SiC in hydrothermal environment [8,9], are being debated. One hypothesis for how silicon species is dissolved in high-temperature water is a mechanism similar to that underlying oxidation in water vapor or steam and it involves formation and subsequent dissolution of a silica film [8,9]. Due to the small activation energy for silica dissolution, once the silica forms, it can quickly dissolve into the hydrothermal environment. This explanation was based on the interpretation of the measurement of X-ray diffraction/photoelectron spectra (XRD/XPS), where Si-O signals were not detected on the corroded surface [8,9]. Nevertheless, it is also recognized that the chemistry of typical water vapor and steam environments are different from the conditions encountered in the hydrothermal environment. In particular, while the high oxygen content in vapor may support the formation of an oxide layer [13,14], the dissolved oxygen concentration in hydrothermal environment is negligible (less than ~5 ppb) [8–11]. Consequently, it is not clear whether the silica layer actually forms under the hydrothermal conditions. With that in mind, another hypothesis has been recently proposed in the literature that involves a direct dissolution of SiC into water via the formation of soluble silicic acid, such as $Si(OH)_4$ [11,15]. While compelling, as of now experimental studies have not provided direct evidence for this hypothesis and the mechanisms underlying transformation of SiC surface into soluble silicic acids have not been proposed or identified. Thus one important question to be yet addressed is whether the silica film forms on the surface of SiC and then quickly dissolves into water or whether there is a direct dissolution process of SiC through the formation of silicic acid on the surface?

In this work we use atomistic simulations based on *ab initio* methods to determine the elementary interfacial reactions that drive hydrothermal corrosion of SiC. Our results include both the thermodynamic analysis of the corrosion products and in-depth understanding of the kinetics based on atomistic mechanisms identified in our simulations.

## 2. Methodology:

### *2.1 Computational details*

Corrosion of water/SiC surface was studied using *ab initio* molecular dynamics (AIMD) within the density functional theory (DFT) framework. We used a symmetric slab (13 atomic layers, 16 atoms per layer, periodically repeated along the (001) direction), with a (2×1) surface supercell periodicity as a model surface. A 16 Å – thick vacuum layer was added to avoid interactions between the periodic images of the slabs. After preparing the SiC surface, liquid water was added to the vacuum region, and it contained 20 water molecules (with the corresponding density of ~0.24 g/cm$^3$). Positions of H$_2$O molecules were identified as discussed elsewhere [16]. In the simulations, we used the canonical ensemble (NVT) with the Nose-Hoover thermostat with a time step of 1 fs. Kinetics of corrosion was simulated at 1000 K. The higher temperature was used to accelerate the surface reactions so that they could be observed within the time scales of AIMD. Here it should be noted that although the water state (~0.24 g/cm$^3$ at 1000 K, ~90 MPa) in our simulation is different the experimental condition (~0.65-0.73 g/cm$^3$ at 548-633 K, ~16 MPa), previous experimental studies have suggested that the influence of water state on the SiC hydrothermal corrosion mechanism (and associated activation energy) is minor [17,18]. We have also run AIMD simulations with a water density of 0.38 g/cm$^3$ at 1000 K and observed similar results. Together these results suggest that our simulation results can be used to understand the mechanisms of hydrothermal corrosion under the experimental condition. All the simulations were performed using the Vienna Ab Initio Simulation Package (VASP) [19]. Settings of the atomistic simulations are discussed below.

Projector-augmented-wave (PAW) potential [20] was used to mimic the ionic cores, whereas the generalized gradient approximation (GGA) in the Perdew-Burke-Ernzerhof (PBE) [21] approach was employed for the exchange and correlation functional. Similarly to our earlier studies, the plane wave cutoff energy was 500 eV, and spin-polarization and dispersion effects using semiempirical corrections (vdW-D) as proposed by Grimme *et al.* [22] were included in the simulations. The integration over the Brillouin Zone was performed with the $\Gamma$ point in the AIMD simulations. To identify the reactants and the activation energy for the elementary corrosion steps, DFT calculations were performed after discovering the elementary reaction pathways in the AIMD simulations. In the DFT calculations, the Brillouin Zone integration was performed by using the 3×3×1 Monkhorst-Pack k-point sampling. The lowest three atomic layers were fixed in the bulk configuration with the bulk SiC lattice constant of 4.378 Å [23]. Relaxation was carried out until forces on ions were lower than 0.01 eV/ Å. Explicit solvent effects were considered in the calculations. Specifically, for all reactions, we considered the possible involvement of a single-water molecule in the explicit water-solvated model [24]. In our calculations we found that the single-water molecule approach is sufficient to describe the explicit effect of water molecules. For each reactive site, a number of siloxane groups are added to cover the available surface around the reactive site, in order to simulate the high coverage of siloxanes on SiC surface in water environment, as observed in the AIMD simulations. Meanwhile, as reported in our previous DFT calculations [16], the species on the surface react locally, thus it is reasonable to assume that the few siloxanes around reactive site is enough to describe the surface reaction. In addition, zero-point energy (ZPE) and entropy corrections were determined, assuming that the adsorbates are bonded strongly enough to have negligible translational and rotational motion, as discussed elsewhere [16]. Activation energies were calculated using the climbing-image nudged elastic band method (CI-NEB) [25]. Linear interpolation was used to generate 7 images for optimization.

The entropy effect on the activation energy has been corrected using the following equation

$$\Delta G_a \sim \Delta F_a = \Delta E_a - k_B T \ln\left(\frac{Z^\neq}{Z^I}\right) \quad (1)$$

where $\Delta E_a$ is the energy barrier calculated from the CI-NEB method after ZPE corrections, $k_b$ is the Boltzmann constant, $T$ is the temperature. $Z^\neq$ and $Z^I$ are the partition functions of an $N$-atom system at the saddle state and the initial state, respectively. The ratio of $\left(\frac{Z^\neq}{Z^I}\right)$ can be written down as

$$\frac{Z^\neq}{Z^I} = \prod_i^{3N \text{ or } 3N-1} \frac{1 - \exp(-\hbar\omega_i^\neq/k_B T)}{1 - \exp(-\hbar\omega_i^I/k_B T)} \quad (2)$$

Here, $\hbar$ is the Planck constant, $\omega_i^\neq$ and $\omega_i^I$ are the vibrational frequencies of the $i^{th}$ mode in the saddle and the initial state, respectively. For the saddle state, only $3N$-1 modes are considered because there is one imaginary vibrational mode along the reaction coordinate.

In addition to the AIMD simulations, additional classical MD simulations were carried out in parallel to search for candidate reaction pathways. The reactive force field ReaxFF [26] was used, as implemented in the Large-scale Atomic/Molecular Massively Parallel Simulator (LAMMPS) package [27]. The detailed settings of MD simulations are the same as in Ref. [16].

## *2.2 Free energy of production of SiO₂ and Si(OH)₄*

To identify the corrosion products on SiC surface, we estimated the formation free energy of $SiO_2$ and $Si(OH)_4$ in a hydrothermal environment. In the case of formation of soluble silicic acid, we have followed a computational approach outlined in our earlier study on corrosion of SiC in molten salts [12]. Here the anodic half-cell reaction can be written as $Si^{SiC} + 4H_2O \leftrightarrow Si(OH)_4^{aq} + 4H^+ + 4e^-$. Assuming the cathodic reaction in water has the form $2H^+ + 2e^- \leftrightarrow H_2$ then the full reaction for the formation of $Si(OH)_4$ can be written as $Si^{SiC} + 4H_2O \leftrightarrow Si(OH)_4^{aq} + 2H_2^{gas}$. The reaction free energy $\Delta G$ of the full reaction can be written as:

$$\Delta G\left(Si(OH)_4^{aq}\right) = \langle E\left(Si(OH)_4^{aq}\right)\rangle - \langle E(H_2O)\rangle - \mu(Si^{SiC}) + 2\mu(H_2^{gas}) - 4\mu(H_2O) + \Delta C_S(Si(OH)_4) + \Delta C_S(H_2O) \quad (3)$$

Using the same cathodic reaction of $2H^+ + 2e^- \leftrightarrow H_2$, the reaction free energy for the formation of solid silica layer, i.e., $Si^{SiC} + 2H_2O^{liquid} \leftrightarrow SiO_2^{solid} + 2H_2^{gas}$, can be written as

$$\Delta G(SiO_2^{solid}) = \mu(SiO_2^{solid}) - \mu(Si^{SiC}) + 2\mu(H_2^{gas}) - 2\mu(H_2O) + \Delta C_S(H_2O) \quad (4)$$

Here, $\langle E(Si(OH)_4^{aq})\rangle$ is the ensemble average of the total internal energy of the dissolved silicic acid in water. $\langle E(H_2O)\rangle$ is the ensemble average of the total internal energy of pure water. $\mu(SiO_2^{solid})$, $\mu(Si^{SiC})$, $\mu(H_2^{gas})$, and $\mu(H_2O)$ are the chemical potentials of solid $SiO_2$, Si species in SiC, hydrogen gas, and an $H_2O$ molecule in water at the temperature and pressure of the calculation, respectively. $\Delta C_S$ is the entropy correction term, describing the change in entropy of one molecule from its standard state to the dissolved one [12]. For example, for the dissolved $Si(OH)_4$ molecule, the magnitude of $\Delta C_S$ is assumed to be the entropy of $Si(OH)_4$ in the ideal gas state [28]. Here the molarity of dissolved $Si(OH)_4$ in the term of $\Delta C_S$ is assumed to be 1 mol/liter (1 M) of $H_2O$. For other concentrations, additional entropy correction for the dissolution of Si species in water is included based on the approximation of dilute species behavior (Henry's law) as $\Delta C_S(C_{Si(OH)4}^{aq}) = \Delta C_S(1\text{ M}) + k_B \ln(C_{Si(OH)4}^{aq})$. For the dissolved $H_2O$ molecule, the magnitude of $\Delta C_S$ is assumed to be the entropy of liquid $H_2O$ at the corresponding target pressure (See SI Sec. 3 and Table S2 for details and values in Supplementary Materials).

According to Eq. (3) and Eq. (4), the formation energy difference between $SiO_2^{solid}$ and $Si(OH)_4^{aq}$ can be written as

$$\Delta G(Si(OH)_4^{aq}) - \Delta G(SiO_2^{solid})$$
$$= \langle E(Si(OH)_4^{aq}) \rangle - \langle E(H_2O) \rangle - 2\mu(H_2O) - \mu(SiO_2^{solid}) + \Delta C_S(C_{Si(OH)4}^{aq}) \quad (5)$$

This equation (Eq. (5)) is also used to describe the reaction free energy for the dissolution of $SiO_2^{solid}$ into water environment as $Si(OH)_4^{aq}$, i.e., $SiO_2^{solid} + 2H_2O \leftrightarrow Si(OH)_4^{aq}$.

The ensemble average of the total internal energy, $\langle E(Si(OH)_4^{aq}) \rangle$ and $\langle E(H_2O) \rangle$, can be obtained through the AIMD simulations at different temperatures, as described in Ref. [12]. As discussed in SI Sec.3, the influence of pressure deviation in the AIMD simulations on $\langle E(Si(OH)_4^{aq}) \rangle$ and $\langle E(H_2O) \rangle$ can be negligible. The temperature effect on the chemical potentials of the relevant solid phases (SiC and $SiO_2$) was included as follows

$$\mu^{solid}(T) = \mu^{solid}(0\text{ K}) + [H(T) - H(0\text{ K})] - T[S(T) - S(0\text{ K})] \quad (6)$$

$\mu^{solid}(0\text{ K})$ is the reference state of the solid from the DFT calculations, $H$ and $S$ are the enthalpy and entropy, respectively, which are tabulated in the JANAF database [29]. The chemical potential of Si species in SiC is related to the chemical potential of the SiC compound and it is constrained within the Si- and C-rich conditions, and the chemical potential of solid $SiO_2$ corresponds to the quartz phase. Similar to the approach outlined in Ref. [30], the chemical potential of hydrogen gas was determined by referencing the JANAF database and the partial pressure of the reference state was assumed to be 1 atm. From Eq. (5), one can see that the variations of hydrogen gas partial pressure and the SiC stoichiometry do not affect the energy difference between $SiO_2^{solid}$ and $Si(OH)_4^{aq}$. The partial molar enthalpy contribution to the chemical potential of $H_2O$ was estimated at different temperature to be $\sim (\langle E(H_2O)_N \rangle / N)$. Here, $N$ is the number of $H_2O$ molecules in the pure water environment. The additional entropy correction for the water is included in the term of $\Delta C_S$. A pressure correction term is also included but negligible, as shown in Table S2 in Supplementary Materials.

## 3. Results and Discussion:

### 3.1. Thermodynamics of corrosion products

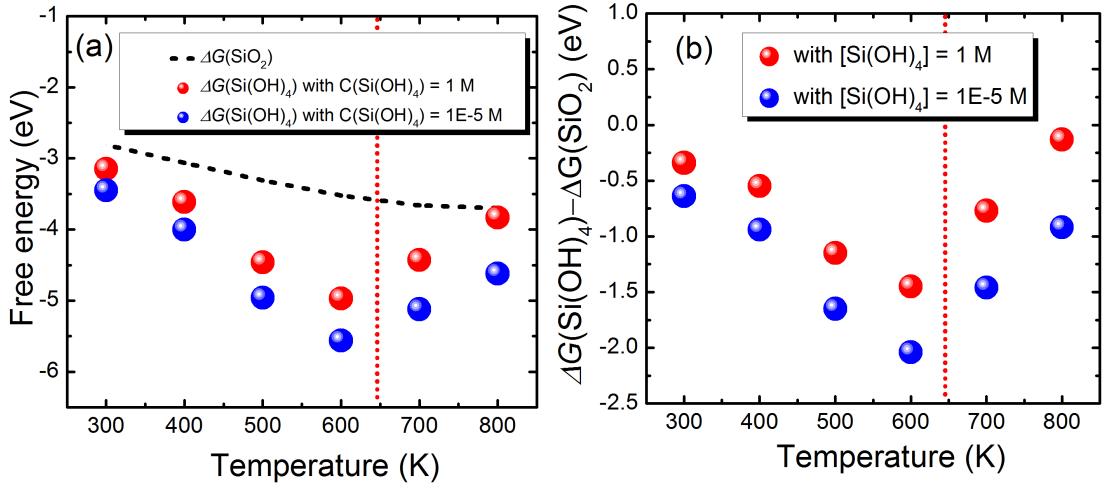

*Fig. 1. Estimated free energies (a) and energy difference (b) for the formation of solid silica and soluble silicic acid during the corrosion of SiC in water at different temperatures. The red dash line corresponds to the water critical point.*

To determine which corrosion products are expected to form on SiC surface, we have calculated the free energy of the formation of the soluble silicic acid ($Si(OH)_4^{aq}$) and of the silica solid layer ($SiO_2$). The results are shown in Fig. 1. We found that, in general, the formation of dissolved silicic acid is more energetically favorable than that of solid silica on the surface. These results indicate that thermodynamically, the hydrothermal corrosion of SiC would eventually form the soluble silicic, regardless of the SiC stoichiometry. This calculation result is consistent with the previous studies of the silica dissolution [31,32]. It is also consistent with the experimental reports of lack of silica on corroded SiC surface, although this thermodynamic calculation does not answer the question of whether solid silica forms on the surface before it is dissolved or whether SiC is directly dissolved as silicic acid. Fig. 1 also shows that, as expected, as the concentration of dissolved Si species decreases (blue symbols), the formation of soluble silicic acid becomes more favorable. Similar results have been reported in earlier studies on the oxidation of single crystal silicon in water at room temperature [33].

As the temperature increases up to the water critical point (~647 K, 22.06 MPa), the energy difference between the formation of $SiO_2$ and $Si(OH)_4^{aq}$ becomes significant, suggesting that there would be a strong preference to form soluble silicic acid and not solid silica on the surface. The increasing stability of silicic acid *vs.* $SiO_2$ in this temperature range is due to the increases in the solubility of silicic acid, e.g., the solubility of silicic acid, ($\ln(C_{Si(OH)4}^{aq})$), in water at 400 K is around 1.22 times of the solubility at 300 K, which is consistent with previous studies showing that the solubility of silicic acid in water at 400 K is around 1.29-1.32 times of the solubility at 300 K [32]. The significantly great stability of $Si(OH)_4^{aq}$ compared to $SiO_2$ at temperatures around 600K may explain why there was no observed solid silica layer on the SiC surface in previous hydrothermal SiC corrosion experiments, which were carried out at ~600 K. The strong preference for the formation of $Si(OH)_4^{aq}$ also suggests that the formation of a solid silica layer may not be a necessary step in SiC corrosion. On the other hand, it is worth noting that as temperature increases above 700 K, the water density becomes significantly reduced and the solubility of $Si(OH)_4^{aq}$ is also reduced significantly [32]. As a result, some of the dissolved species could precipitate out as solid silica on the surface, especially at high concentration of dissolved Si species in the solution.

### *3.2. Proposed SiC hydrothermal corrosion mechanisms*

In Fig. 2 we show a schematic of the possible reaction paths that were identified for SiC dissolution in hydrothermal condition. After $H_2O$ chemisorption, the process of SiC corrosion can be divided into three stages: (i) Surface attack, which involves breaking of Si-C surface bonds through the hydrogen scission; (ii) Surface evolution, which is characterized by formation and transformation of surface motifs/species; (iii) Surface dissolution, which is characterized by dissolution of Si species into the surrounding water. Hydrogen scission was identified in Ref. [16], based on DFT calculations with an implicit solvent model. In Section 3.3, we will discuss the effect of the explicit solvent on the activation energy of hydrogen scission reactions. Surface evolution and dissolution steps after the initial hydrogen scissions will be discussed in Sections 3.4 and 3.5, respectively, and we summarize the key results here. After the initial H scission reaction, the displaced Si atoms on the surface are either a $H_2SiO_3$ motif or a $SiO_3(OH)$ motif (see Fig. 2). The major branching point for determining which motifs are produced is based on whether hydrolysis reactions occur after the hydrogen scission. We describe this differentiation as progressing through two paths: Path I for the formation of a $H_2SiO_3$ motif, which was observed in the AIMD simulations, and Path II for the formation of a $SiO_3(OH)$ motif, observed in the classical MD simulations. Regardless of which of the two characteristic motifs are produced, both of them become hydrolyzed and eventually dissolve into water as $Si(OH)_4$. Interestingly, the dissolved $Si(OH)_4$ species has been previously reported to be a product of silica dissolved in water, through a set of hydrolysis and deprotonation reactions in [34–38], similar to those found by us on SiC surface. However, in the current work, our kinetic studies reveal that instead of first forming silica layer above the surface, $Si(OH)_4$ can be directly produced on the SiC surface. The elementary steps shown in Fig. 2 and described briefly above will be discussed in detail in subsequent sections. Our calculations were carried out using explicit models for water to gain a better understanding of the impact of the solvation and spatial freedom of water within our DFT calculations. In Section 3.6, we will compare the DFT identified reaction rate with the experimentally measured corrosion rate, and explore the rate-limiting step during the dissolution of Si species in SiC. Finally, in Section 3.7 we will describe the effects of the crystallographic surface orientation on SiC hydrothermal corrosion, and we will show that the hydrogen scission reactions play a key role in surface degradation, regardless of the surface orientation.

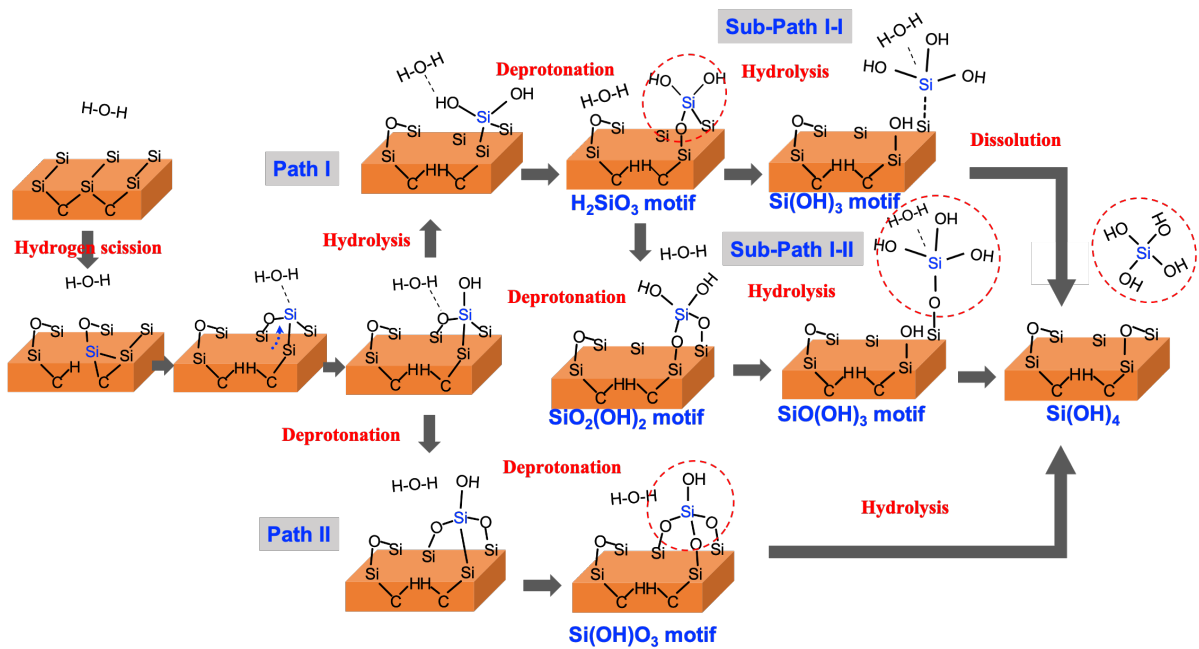

*Fig. 2. A schematic showing reaction paths for SiC hydrothermal corrosion identified in our simulations. The blue Si atoms are those that have been displaced form SiC surface.*

### 3.3. Hydrogen scission reaction in the explicit solvent model

Hydrogen scission has been previously shown to be an important mechanism for the initial attack of water on SiC surface [16]. In this process, the hydrogen atom emitted from the bridging hydroxyl group on the surface breaks the Si-C bonds and thus attacks the surface. While Ref. [16] used an implicit water model, a solvated water model, which involves a single-water molecule in the reaction step, has been explicitly considered in the current work to identify the solvent effect on the hydrogen scission reactions. In addition to the interaction between solvent water molecules and the hydroxyls on the surface, the spatial freedom of these water molecules has also been considered by sampling the water molecules at different spatial positions. The sampling was carried out in the AIMD simulations at low temperature (e.g., ~300 K) to avoid the dissociation of solvent water molecules. The AIMD simulations were followed by DFT calculations to determine the activation energy of the hydrogen scission reaction.

The structures of the initial, transition, and final states from the explicit-solvent-model calculations are shown in Fig. 3. The initial and final structures were determined in the AIMD simulations, which were optimized by the following DFT calculations, and the transition structure was obtained in the CI-NEB calculations. Depending on the spatial position of water molecule, the reaction enthalpies for the hydrogen scission reactions range from -0.23 to -0.05 eV, which are somewhat more positive than those calculated in the implicit model, i.e., from -0.74 to -0.21 eV [16]. These results suggest that the existence of explicit solvation water slightly increases the reaction energy (*i.e.* destabilized products *vs.* reactants and make the reaction enthalpy more positive) of the hydrogen scission reaction. The reaction activation energy was found to be substantially increased after the explicit solvent effect is included. The enthalpy ranges from 1.44 to 1.82 eV (the corresponding activation free energy at 600 K ranges from 1.42 to 1.78 eV), whereas the average barrier without the explicit solvation water is ~1.16 eV. The difference in the activation energy with and without explicit single-water solvating is due to the hydrogen bonding interaction between $H_2O$ molecules and the hydrogen atom emitted from surface hydroxyl groups (see the dashed line in the transition state in Fig. 3). This hydrogen bonding interaction between the solvating $H_2O$ molecule and the emitted H atom competes with the electrostatic interaction between the emitted H atom and C subsurface atom, and thus impede the hydrogen scission reaction. To verify that a single molecule is sufficient to include the explicit solvent effect, we introduced one more water molecule, and the activation barrier with two solvation water involved was calculated to be ~1.47 eV, suggesting that the explicit solvent effect on hydrogen scission reaction does not change as the number of solvation water increases.

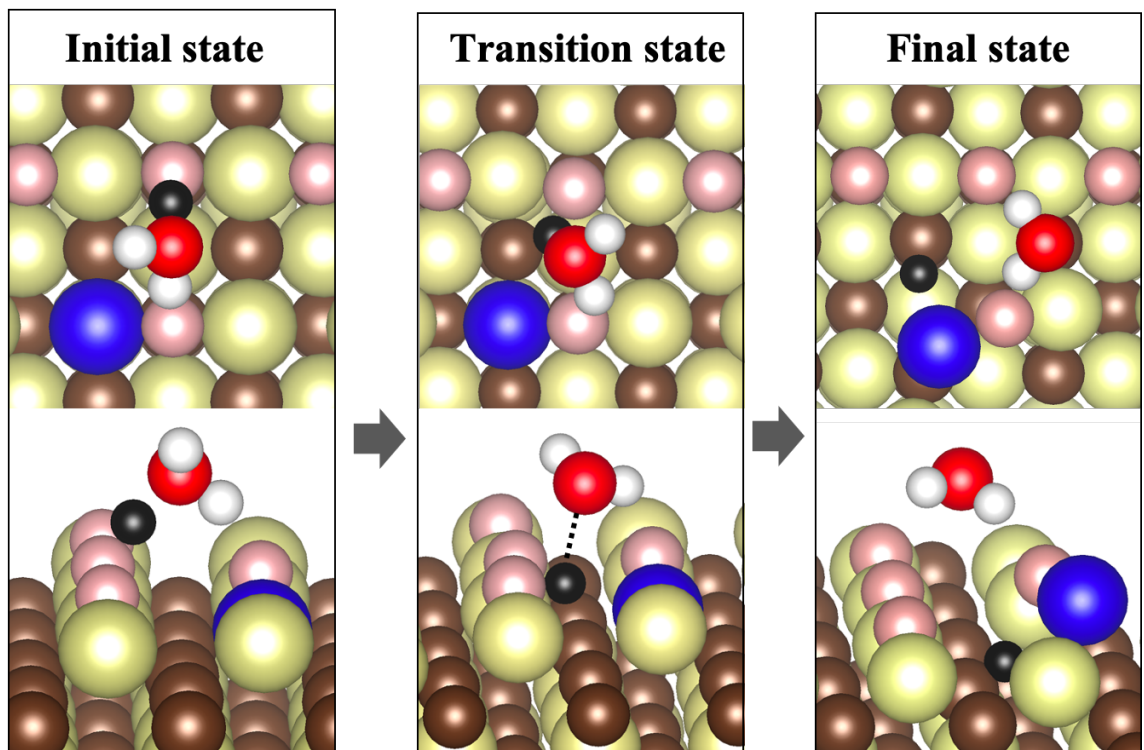

*Fig. 3. Optimized DFT structures of the initial, transition, and final states associated with the hydrogen scission reaction in the explicit water model. The initial state shows the existence of solvating $H_2O$ molecule around the formed bridging hydroxyl group on the surface; the transition state shows the hydrogen bonding interaction between the emitted H atom from the bridging hydrolysis group and the solvating $H_2O$ molecule; and the final state shows the emitted H atom terminates the subsurface C atom and displaces the surface Si atom. The yellow spheres are the Si atoms, and brown ones are the C atoms, the light red spheres are the adsorbed O atoms on the surface, and the dark red spheres are the O atoms in $H_2O$ molecules, and white spheres are the H atoms. The black sphere is the emitted H atom, and the blue sphere denotes the displaced Si atom.*

### 3.4. Formation of characteristic motifs

Once the hydrogen scission reactions occurred on the surface, the surface Si atom is displaced from its surface site, but still bound to the oxygen atom on the surface in the form of *Si-O-Si linkage (here, *Si denotes the surface Si atom to distinguish it from the displaced Si atom). Subsequently, the displaced Si atom and its surrounding water molecules interact via multiple hydrolysis and deprotonation reactions, leading to formation of either $H_2SiO_3$ (Path I) or $SiO_3(OH)$ (Path II), as shown in Fig. 2. Both paths shown in Fig. 2 are possible in principle and have been observed in our atomistic simulations at high temperature. Specifically, alternative hydrolysis and deprotonation reactions occurred along Path I, which resulted in the production of $H_2SiO_3$, whereas only deprotonations were observed in the formation of $SiO_3(OH)$ along Path II. In the following, we will discuss these two paths in detail and determine which one is energetically favorable.

      The hydrolysis reactions along Path I are similar to those observed during silica dissolution [34–38]. The optimized structures of the initial, the transition, and the final states for the hydrolysis reaction with explicit solvation are shown in Figs. 4(a)-(c). Specifically, in this process one $H_2O$ molecule around the displaced Si atom dissociates into -H and -OH groups. The -H binds to a bridging oxygen atom and cleaves the Si-O bond, while the -OH binds to the surface *Si atom (*i.e.*, HO-Si-O-Si* + $H_2O$ → $Si(OH)_2$ + *Si-OH, see Fig. 4(c)). One can see that, at the transition state (Fig. 4(b)), the hydrogen atom in $H_2O$ weakly interacts with the bridging oxygen atom, and the bond length between the surface Si and the bridging oxygen is extended from 1.70 to 1.75 Å, leading eventually to breakage of the *Si-O bond. The reaction enthalpy for the aforementioned reaction ranges from -1.38 to -0.77 eV, suggesting this reaction is energetically favorable. The energy barrier for this reaction ranges from 0.40 to 0.87 eV, which is consistent with the previously reported barriers for hydrolysis reactions in silica dissolution, ~0.62-0.99 eV [34–38].

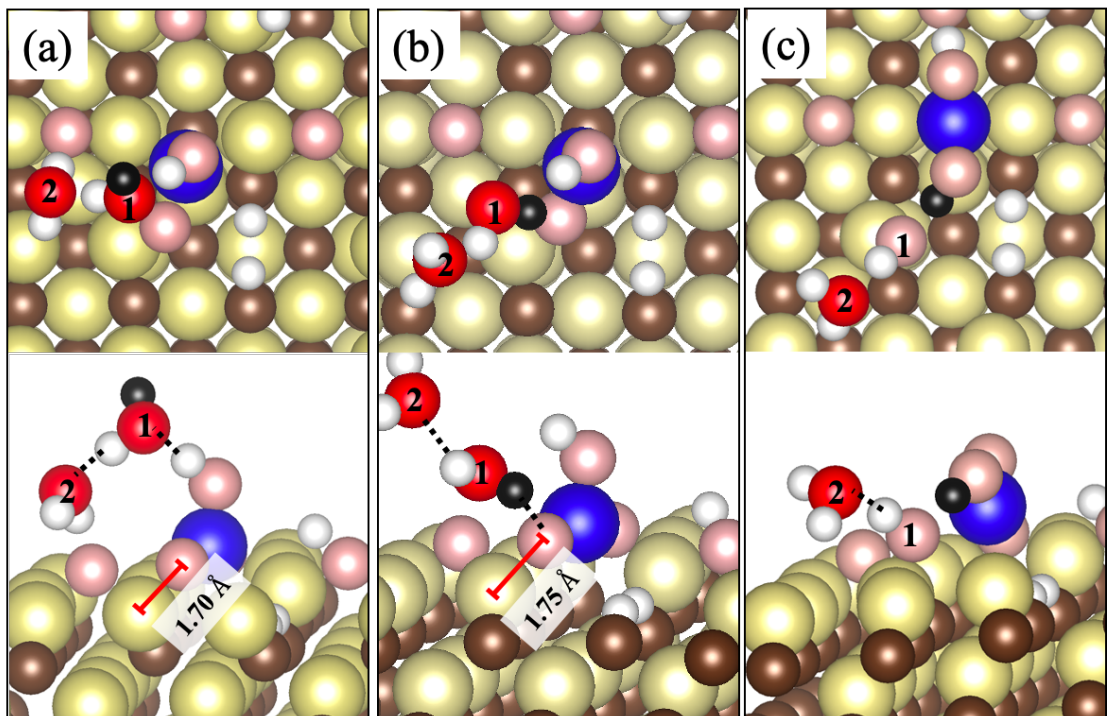

*Fig. 4. Optimized DFT structures associated with the hydrolysis reaction, e.g., HO-Si-O-Si\* + $H_2O$ → $Si(OH)_2$ + \*Si-OH in Path I. (a)-(c) are the initial, transition, and final states for the hydrolysis reaction, respectively. (a) The initial state shows the existence of $H_2O$ molecules around the previously formed HO-Si-O-Si\*; (b) the transition state shows the interaction of H atom in $H_2O$ molecule with the bridging O atom in HO-Si-O-Si\*, which tends to elongate the distance between surface \*Si atom and the bridging O atom; and (c) the final state shows the dissociation of $H_2O$ molecule into -H and -OH; the -H binds with the bridging O atom to form $Si(OH)_2$, and the -OH binds with the surface Si\* atom. The yellow spheres are the Si atoms, and brown ones are the C atoms, the light red spheres are the adsorbed O atoms on the surface, and the dark red spheres are the O atoms in $H_2O$ molecules. The white spheres are the H atoms. The blue sphere denotes the displaced Si atom.*

Because of the occurrence of the subsequent deprotonation reaction in Path I, the newly formed $Si(OH)_2$ loses a hydrogen atom from one of the two hydroxyl groups through the proton exchange mechanism, i.e., $Si(OH)_2$ + $H_2O$ → Si(OH)-$O^-$ + $H_3O^+$ → -O-$Si(OH)_2$ + 2\*H (see Fig. 5(a)-(c)). At the transition state, the hydrogen atom from a hydroxyl group transfers to a nearby $H_2O$ molecule, forms $H_3O^+$, and leaves behind the undercoordinated oxygen on the surface (as indicated by an arrow in Fig. 5(b)). In the following processes, one can see that the undercoordinated oxygen species will interact with its nearby surface Si atoms by subsequently rotating and inserting between the displaced Si and the surface Si atom, as shown in Fig. 5(b). Such deprotonation-induced oxygen-incorporation reaction has been previously observed in the oxidation process of single crystal silicon [39]. Afterwards, the undercoordinated Si atom will attract additional $H_2O$ molecule (see Fig. 5(c)), and eventually form $H_2SiO_3$. The barrier for deprotonation along Path I falls in the range between 1.31 and 1.40 eV at 0 K, which is comparable to the deprotonation barriers along Path II (ranging from 1.32 to 1.44 eV). When considering the temperature effect, the barrier for the deprotonation along Path I is reduced and at 600K, it falls in the range of 1.05 and 1.27 eV (1.08-1.31 eV for the deprotonation along Path II). In addition, it is worth noting that the activation energy for deprotonation reaction is much higher than that for hydrolysis reaction (0.40-0.87 eV, see Fig. 6), indicating that the reaction rate along Path I should be significantly faster as compared with that along Path II (see Fig. 2). As a result, it is expected that the formation of $H_2SiO_3$ motif would be dominant in the process of hydrothermal corrosion, and in the following discussion we will mainly focus on the reactions in Path I.

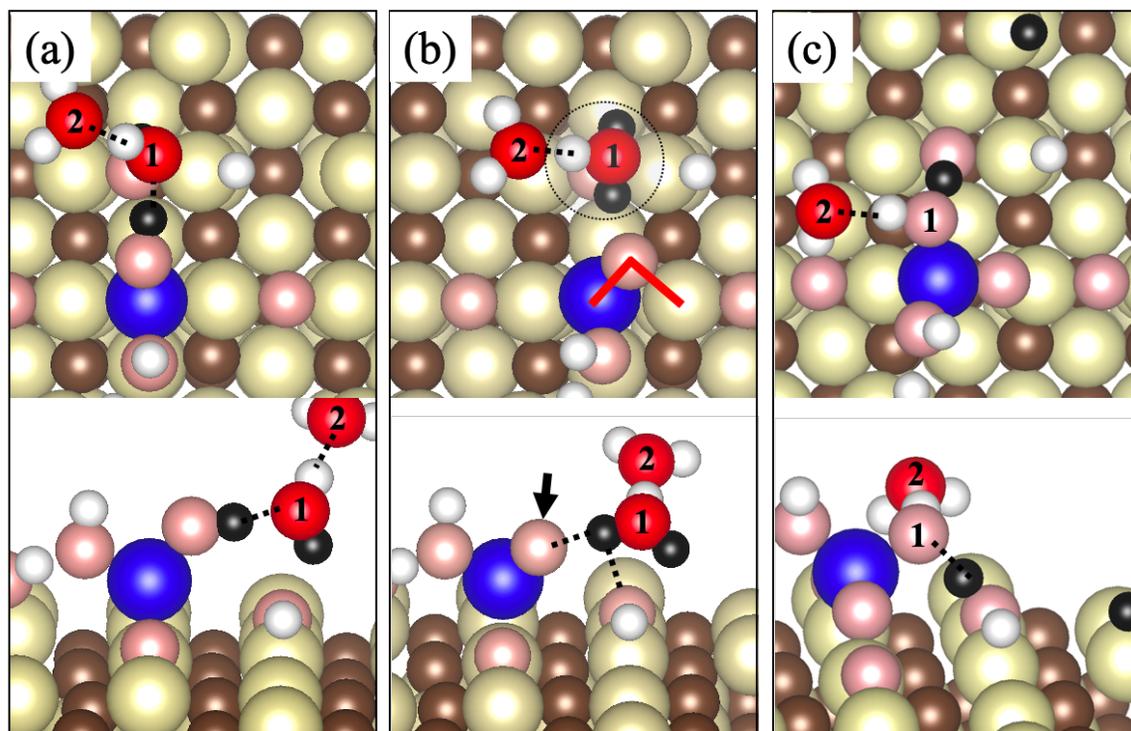

Fig. 5. Optimized DFT structures associated with the deprotonation reaction, e.g., $Si(OH)_2$ + $H_2O$ → $Si(OH)$-$O^-$ + $H_3O^+$ → -O-$Si(OH)_2$ + 2*H in Path I. (a)-(c) are the initial, transition, and final states for the hydrolysis reaction, respectively. (a) The initial state shows the existence of $H_2O$ molecules around the formed $Si(OH)_2$; (b) the transition state shows the formation of $H_3O^+$, which leaves one undercoordinated oxygen, as arrowed in figure, on the surface; and (c) the final state shows the proton transition through $H_3O^+$, and also the interaction between the undercoordinated displaced Si atom and the $H_2O$ molecule, which finally forms the $H_2SiO_3$ motif. The yellow spheres are the Si atoms, and brown ones are the C atoms, the light red spheres are the adsorbed O atoms on the surface, and the dark red spheres are the O atoms in $H_2O$ molecules, and white spheres are the H atoms. The blue sphere denotes the displaced Si atom.

## 3.5. Evolution and dissolution of characteristic motifs

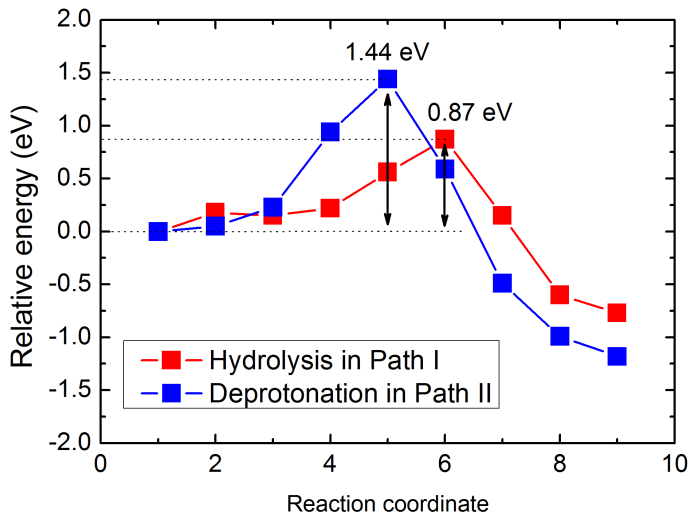

*Fig. 6. DFT calculated energies barriers for the hydrolysis reaction (0.40-0.87 eV) along Path I and deprotonation reaction (1.32-1.44 eV) along Path II. For clarification, only the top range of these barriers are plotted.*

Once the $H_2SiO_3$ motif in Path I was produced on the surface, it was stable within the AIMD simulation time scales (~100 ps). However, earlier DFT studies on silica dissolution have shown that the $H_2SiO_3$ motif is an intermediate state and that it will continue to be hydrolyzed by the surrounding water molecules [40]. The difference between the published results and our simulations maybe in part due to the strong interaction between water molecules and the $H_2SiO_3$ motif introduced in Ref. [40], where the authors artificially brought the $H_2O$ molecule in close proximity to the $H_2SiO_3$ motif, so to drive a strong attack of water on the $H_2SiO_3$ motif. Our DFT calculated reaction barrier for the hydrolysis of $H_2SiO_3$ motif (e.g., *Si-O-Si(OH)$_2$ + $H_2O$ → *Si-OH + Si(OH)$_3$), ranges from 0.75 to 0.96 eV, suggesting that once the water molecule has approached the $H_2SiO_3$ motif, it can be easily hydrolyzed. Another reason for lack of direct observation of $H_2SiO_3$ hydrolysis in our AIMD simulations might be that, once $H_2O$ molecules were adsorbed onto the surface in our simulations, the water density above the surface was reduced, weakening the interaction between water and the $H_2SiO_3$ motif. The lower water density could to some extent reduce the probability of the hydrolysis reaction, and it may require a longer simulation time to occur. However, due to the low hydrolysis reaction activation energy, it is expected that the $H_2SiO_3$ motif is an intermediate motif that will eventually evolve to Si(OH)$_3$ through the hydrolysis reaction (see the sub-path I-I in Fig. 2).

After forming Si(OH)$_3$, which is bonded to the surface with a single chemical bond, it is reasonable to expect that a dissolution reaction will take place. As observed in our continuous AIMD simulations, the dissolution of Si is induced by the surrounding water molecules. Specifically, water molecules interact with the displaced Si atom and induce the formation of the penta-coordinated Si, accompanying with the elongation of the *Si-Si bond distance from 2.33 to 2.52 Å, and eventually break the *Si-Si bond (see Fig. 2). The displaced Si is finally dissolved as Si(OH)$_4$. The calculated activation energy for above dissolution reaction was determined to be in the range of 0.62 and 0.69 eV.

Besides the evolution along the Sub-Path I-I, another $H_2SiO_3$ evolution pathway (denoted as the Sub-Path I-II) has been observed in our classical MD simulations. It involves the deprotonation and hydrolysis reactions. The alternative deprotonation/hydrolysis reactions induce the transformation of $H_2SiO_3$ into the $SiO(OH)_3$ configuration, which is eventually dissolved as $Si(OH)_4$ (see Fig. 2). Specifically, similarly to the reactions in the process of $H_2SiO_3$ formation (see Fig. 5 in Sec. 3.4), the deprotonation reaction along the Sub-Path I-II is mediated by the proton-exchange transfer, coming with the oxygen-incorporation between the displaced Si and the surface Si atom. After deprotonation, $H_2SiO_3$ motif transforms to the $SiO_2(OH)_2$ configuration (see Fig. 2). Subsequently, a quick hydrolyzation of $SiO_2(OH)_2$ leads to formation of $SiO(OH)_3$ through cleavage of the Si-O bond. As observed in the MD simulation, the formed $SiO(OH)_3$ motif is unstable and eventually dissolved as $Si(OH)_4$. However, it should be noted that the calculated reaction barrier for the deprotonation reaction along Sub-Path I-II is around 1.34 eV, which is much higher than the barrier for the hydrolysis along Sub-Path I-I. This higher barrier suggests that the reaction rate along Sub-Path I-II would be much slower than that along Sub-Path I-I. Therefore, it is expected that the evolution of $H_2SiO_3$ motif along Sub-Path I-I would be dominant in the process of hydrothermal corrosion.

The above analysis of sub-paths along Path I confirms that the $H_2SiO_3$ motif observed in our simulations is a metastable species and will eventually dissolve as $Si(OH)_4$ in the aqueous environment. More generally, it is interesting to note that although a stable silica layer does not form on the SiC surface, the dissolution reactions and characteristic motifs are similar to those reported during the silica dissolution. These observations are consistent with our thermodynamic analysis presented in Sec. 3.1 and further indicate that the formation of silica layer is not a necessary step during SiC hydrothermal corrosion.

## *3.6. Discussion of the rate-limiting step for Si dissolution*

According to above results, we can see that there are four kinds of reactions during the dissolution of Si species on SiC surface, involving the hydrogen scission, hydrolysis, deprotonation, and the dissolution reaction (see Fig. 2). To determine which reaction is the rate-limiting step, we have summarized the calculated activation energies for the above reactions in Table 1. As discussed in Sec. 3.4, Path I and its Sub-Path I-I are the fast reaction pathways for Si dissolution, therefore we only show the activation energies of the reactions along Path I and Sub-Path I-I. From Table 1, we can see that the hydrogen scission reaction has the highest activation energy along the dissolution path.

Ideally, we would be able to compare reaction activation energies determined from DFT to activation energies determined from dissolution experiments performed at different

*Table 1. The calculated activation energy of reactions during the dissolution rate of Si species along Path I and its Sub-Path I-I.*

| Reaction | Activation energy at 600 K (eV) |
|---|---|
| Hydrogen scission | 1.42-1.78 |
| Hydrolysis | 0.35-0.81 |
| Deprotonation | 1.05-1.27 |
| Dissolution | 0.59-0.62 |

temperatures. However, temperature-dependent experimental dissolution rates at not available.

Instead, to compare to experiments, we use experimental dissolution rates measured at a single temperature and we use a simplified model to provide an approximate estimate of the activation energies (see Table 2). Specifically, for the dissolution of SiC in hydrothermal environment, the experimentally measured dissolution rate, $r$, can be described as

$$r = k_f[\text{SiC}][\text{H}_2\text{O}]^n$$

Table 2. The experimentally measured dissolution rate of SiC in hydrothermal conditions and the corresponding estimated activation energy.

| Condition | Dissolution rate, $r$, [mg/cm$^2$s] | Rate constant, $k_f$, [1/s] | Activation energy, $E_a$, [eV] |
|---|---|---|---|
| 633 K, 18.5 MPa, CVD SiC [a] | $4.05 \times 10^{-7}$ | 0.09 | 1.74 |
| 633 K, 18.5 MPa, CVD SiC [b] | $5.64 \times 10^{-8}$ - $1.53 \times 10^{-6}$ | 0.01-0.35 | 1.67-1.85 |
| 603 K, 15 MPa, CVD SiC [c] | $1.16 \times 10^{-8}$ | 0.003 | 1.84 |

a. Ref [9], b. Ref. [15], c. Ref. [8].

Here, $k_f$ is the reaction rate constant for SiC dissolution, [SiC] and [H$_2$O] are the concentrations of reactive species on SiC surface and H$_2$O at the reaction interface, respectively. $n$ is the reaction order, which depends on the number of water molecules involved in the reaction. At the reaction interface, water is abundantly available, indicating that the reaction rate is independent of the concentration of water molecules, *i.e.*, at zero-order approximation. Consequently, the dissolution rate of SiC in hydrothermal environment is determined by the surface concentration of reactive species and the rate constant. As mentioned above, the initial dissolution of SiC is mainly due to the dissolution of Si species, therefore, the reactive species is assumed to be the concentration of Si species on the surface. For the area of 1×1 cm$^2$ SiC(100) surface, the surface concentration of Si species is ~4.31×10$^{-6}$ mg/cm$^2$. For the dissolution rate constant, we use the transition state theory and we can write an Arrhenius form of dissolution rate constant as, $k_f = C * \exp(-E_a/RT)$. The pre-exponential factor is approximated here as: $C = \tau k_b T / h$, where $R$ is the gas constant, $k_b$ is the Boltzmann constant, $T$ is the absolute temperature, $h$ is Planck's constant, and $\tau$ is the transmission coefficient, which represents the probability that the system will go over the barrier, and normally within 0 and 1. Since the probabilities of the back/forward reactions are equal to each other. A reasonable value of $\tau$=0.5 is used in the calculations. We tested the effect of $\tau$ and found that the activation energy will be increased by only ~0.30 eV, as $\tau$ varies from 0.001 to 1, which suggests that the variation of transmission coefficient will not change the final conclusion. Based on the experimentally measured dissolution rates, the activation energy, $E_a$, for SiC dissolution was estimated to be in the range from 1.67 to 1.85 eV, which is in relatively good agreement with the energy barrier (~1.42 to 1.78 eV) for hydrogen scission reactions determined from our DFT calculations. These results suggest that the hydrogen scission reaction could be an essential step for SiC dissolution. Interestingly, hydrogen breaking of chemical bonds has been previously observed and reported to be an important step in the Smart Cut technology to produce Silicon On Insulator wafers [41] and in the degradation of SiO$_2$-based insulated electronic devices [42].

## 3.7. Corrosion on different surface orientations

It is known that the crystallographic orientation of the surface could potentially have a significant effect on corrosion and dissolution rate [43–46]. To identify this effect during hydrothermal corrosion of SiC, we have performed additional *ab initio* simulations and analysis based on the generalized Butler-Volmer Kinetics model [47]. According to this model, the propensity for corrosion is correlated with the surface energy density, defined as $E_{surf}/\rho$, where $E_{surf}$ is the surface energy of a given surface, and $\rho$ is the surface atomic density. A more detailed discussion of this approach can be found in the Supplementary Materials. Specifically, the higher the surface energy density, the larger the corrosion current density [48].

The surface energy density for different crystallographic orientations of SiC was calculated using DFT and it is plotted as a function of the Si chemical potential in Fig. 7. These calculations show that, under Si-rich conditions, the corrosion current density *I* decreases in the order of $I$[C(111)]> $I$[Step-C(100)]> $I$[C(100)]> $I$[Si(111)]> $I$[Si(100)]> $I$[Step-Si(100)]> $I$[(110)]. The surfaces with steps (Step-Si(100) and Step-C(100)) will be discussed below in more detail. The above order of the corrosion current density means that the (110) surface has the slowest dissolution rate, owing to its lowest surface energy density. Interestingly, this prediction is also in close agreement with our AIMD simulation observations (See Fig. S1 in Supplementary Materials). In Fig. S1, we can see that the (110) surface is orderly passivated by Si-OH and C-H groups, which are stable within our AIMD simulation time scale (~100 ps). In addition, from Fig. 7 we can see that the surface energy densities of the C-terminated surfaces are generally higher than those of Si-terminated surfaces, which suggests that the C-terminated surfaces could have a faster dissolution rate than the Si-terminated ones. Similar results have also been observed during dry thermal oxidation of SiC, where the oxidation growth rate on C-terminated surface is significantly larger than that on Si-terminated surface [49]. These faster corrosion rates of C-terminated surfaces suggest that the rate-limiting step for SiC hydrothermal corrosion occurs on the Si-terminated surfaces. Again, these results are consistent with the good agreement we have shown between the activation energy for surface corrosion on Si-terminated surfaces and published experiments.

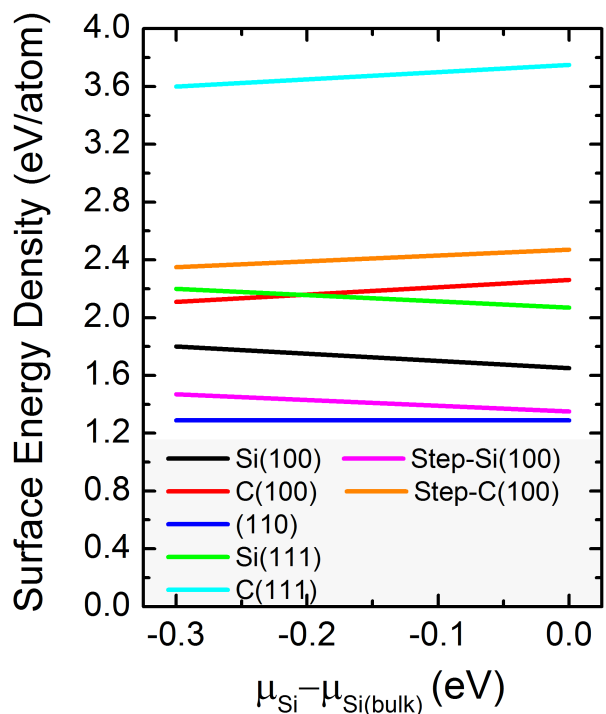

*Fig. 7. Estimated surface energy density for different SiC surface configurations as a function of Si chemical potential.*

To understand the dynamic corrosion processes of C-terminated surface, we have also performed AIMD simulations for the C-terminated surface, as shown in Fig. 8. In Fig. 8(b), we can see that similar to the Si-terminated surface, the hydrogen scission reactions induce the surface attack on the C-terminated surface. The calculated activation energy for the initial hydrogen scission reaction is ~1.30 eV (the corresponding activation energy at 600 K is ~1.26 eV), which is lower than that on the Si-terminated surface (larger than ~1.42 eV at 600 K), as discussed in Sec. 3.3. After the H scission reaction, Si is promptly pulled out from the second layer beneath the surface and binds to -OH group. At the same time, the topmost C atoms tend to swap positions with the second layer Si atoms, resulting in the formation of a carbon ring structures (See Fig. 8(c)). Similar swapping mechanism has been reported in *ab initio* simulations of molten salt corrosion of SiC [12]. The swapping reaction exposes displaced Si atom to the water environment, and these Si atoms are further hydrolyzed and deprotonated to form the $H_2SiO_3$ motif on the surface (see Fig. 8(d)). Due to the low activation energies for its subsequent evolution and dissolution (see Sec. 3.4), it is expected that the $H_2SiO_3$ motif would rapidly dissolve into the water environment. After Si dissolution, the remaining topmost C layer is disordered and forms a ring-like structure (with a high concentration of $sp^2$ bonds, similarly to the carbon-rich layer reported to form during molten salt corrosion [12]). The newly formed carbon structures could be the precursors for the experimentally observed carbon-rich layers in the early stage of hydrothermal corrosion [8–11]. Our simulations confirm the results of earlier study that hydrogen scission reactions play an important role in SiC hydrothermal corrosion, both on C- and on Si- terminated surfaces.

We have also considered the effect of surface steps on reaction rates. Such steps have been reported for example on the (100) surface of SiC grown on Si substrate [50–52]. To determine the role of these steps, we have also calculated the surface energy density for the (100) surfaces that contain steps. The step structure was built on the basis of experimental observations [50–52], where the step has {111} facets. Here, we focus on the mono-bilayer steps [50] since the cliff-like multi-layer steps have higher formation energies, owing to the difficulty in forming stable reconstruction among step edge atoms. From Fig. 5, we can see that for the C-terminated surfaces, the surface energy density for the step surface is slightly larger than that of the flat surface, suggesting that on C-terminated surfaces, the step could to some extent accelerate chemical degradation of the surface. This trend may be due to a larger open space present near the step than on flat C-terminated surface – this open space increases the exposure of the second layer Si to $H_2O$ molecules. Interestingly, for the Si-terminated surfaces, the surface energy density for the step surface is smaller than that of the flat surface (although the surface energy for the step surface is larger than that of flat surface), suggesting that the Si-terminated step surface would a have slower dissolution rate as compared to the flat one. To explore the mechanism for this slower dissolution rate, we have performed additional AIMD simulations to detect reaction mechanisms and to determine their kinetics. Similarly to other surfaces, the hydrogen scission attack mechanism is observed in the AIMD simulations (within the first ~50 ps). However, our DFT calculations show that the reaction enthalpy for this H scission reaction ranges from 0.17 to 0.53 eV. The positive reaction energy suggests that the hydrogen scission reaction would be energetically less favorable on the Si-terminated step surface as compared with the Si-terminated flat surface. Similar phenomena have been reported during oxidation of single crystal silicon [53], where it was found that the oxidation of silicon is a terrace-attacking process and the surface steps do not affect the oxidation rate [53].

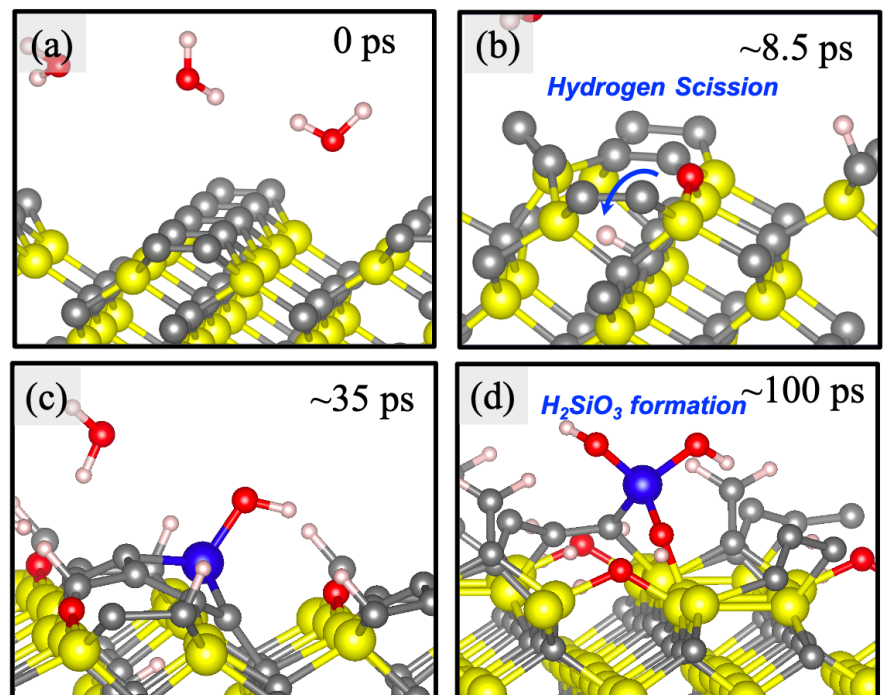

*Fig. 8. Snapshots from an AIMD simulation at 1000 K for the corrosion evolution on the C-terminated surface. The yellow spheres are the Si atoms, gray ones are the C atoms, red ones are the O atoms, and white ones are the H atoms. The blue sphere represents the displaced Si atom after hydrogen scission reaction.*

## 4. Conclusion:

We have identified atomic-level mechanisms driving chemical degradation of SiC by high-temperature water and we have quantified the activation energy barriers of the corresponding interfacial reactions. This study provides significant new understanding of the hydrothermal corrosion process beyond the earlier studies of water chemisorption on surfaces. The initial chemical attack of water on the surface is facilitated by a hydrogen scission reaction, regardless of the surface orientation and passivation. It was found that hydrogen scission reaction continues to play a key role as corrosion continues, as this is the main mechanism of breaking Si-C bonds. Even for surfaces that are C- terminated, Si is dissolved first after it has segregated to the surface through a number of swapping reactions. These swapping reactions and Si segregation were previously reported during corrosion of SiC in molten salts [12], which means that this might be a common mechanism for corrosion of multi-component ceramics where one species is significantly more reactive than another.

One of the important findings from this study is that silica does not form on the surface, but that SiC is directly dissolved into the aqueous environment. Interestingly, the chemical reactions that facilitate dissolution of SiC surfaces are the same as previously reported for dissolution of Si. The reaction energy barriers are not the same as for dissolution of silica, but they are comparable. The rates for dissolution of SiC determined from the calculated activation energy barriers are consistent with rates estimated from available experimental data on hydrothermal corrosion of SiC.

Si is dissolved by forming intermediate motifs (e.g., $H_2SiO_3$) on the surface, which later are transformed into soluble silicic acid $Si(OH)_4$ through a series of hydrolysis and deprotonation reactions. Once Si is dissolved, it leaves behind a C-rich surface with high $sp^2$ ratio, which is likely a precursor to the experimentally observed C-rich layer. In general, the C-terminated surfaces was found to have a faster dissolution rate than the Si-terminated surface, which is consistent with previous results in the dry thermal oxidation of SiC [49].

Finally, we have also considered different surface crystallographic orientations and stepped surfaces. We found that steps can accelerate key surface reactions on C- terminated surface, but not on Si- terminated surface. We have found that the (110) surface would be the most corrosion-resistant surface as compared to other stable surfaces of SiC, which could guide strategies for the design of high corrosion-resistant SiC for hydrothermal applications.

## Acknowledgement

This research is supported by the U.S. Department of Energy, Nuclear Energy University Program, under Grant No. DE-NE0008781.

# Supplementary Materials:

1. **Corrosion evolution on the (110) surface**

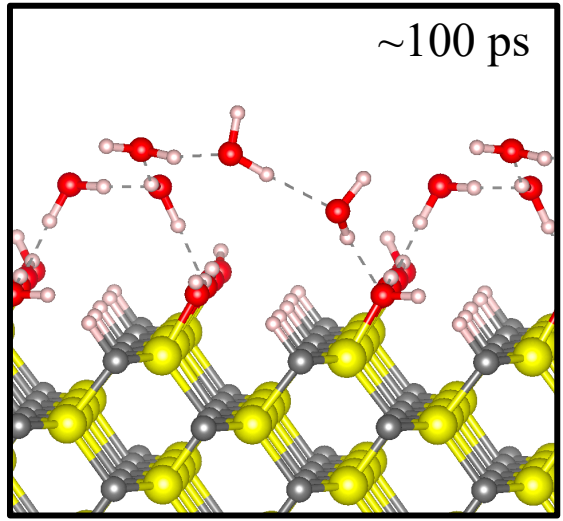

*Fig. S1. Snapshots from an AIMD simulation at 1000 K for the corrosion evolution on the (110) surface. The yellow spheres are the Si atoms, gray ones are the C atoms, red ones are the O atoms, and white ones are the H atoms. The surface is passivated by Si-OH and C-H groups.*

2. **Surface energy calculation**

As discussed in main text, the periodical slab with 16 Å vacuum layer is utilized to represent the surface. For both of the surfaces along (100) and (110) directions, we use the symmetric slabs with 13 atomic layers and 16 atoms per layer. The description of step surface along (100) direction can be obtained in main text, and due to the open space, there are 20 atoms on the step surfaces.

The surface energy for each slab can be written as:

$$\gamma = (E_{slab} - n_{Si}\mu_{Si}^{SiC} - n_C\mu_C^{SiC})/2A$$

Here, $E_{slab}$ is the total energy of slab, $n_{Si\ or\ C}$ are the number of Si or C species within the slab, respectively; $\mu_{Si\ or\ C}^{SiC}$ are the chemical potential of Si or C species in SiC, respectively, and $A$ is the surface area. The calculated surface energies along (100) and (110) directions are listed in Table S1.

The situation of surfaces along (111) direction is more complicated. For the C- and Si-terminated SiC(111) surfaces, both of them have two kinds surfaces depending of the number of dangling bonds of each surface atom: i.e., one-dangling-bond or three-dangling-bonds per each surface atom. Therefore, the SiC(111) surfaces have four different cases: C-terminated with one dangling bond (C1); C-terminated with three dangling bonds (C3); Si-terminated with one dangling bond (Si1); Si-terminated with three dangling bonds (Si3). By using the periodical slab approach, four asymmetric slabs along (111) direction with 8 atoms per layer were built, involving (1) 14-layers with C1 and Si1 terminations, denoted as C1Si1; (2) 14-layers with C3 and Si3 terminations, C3Si3; (3) 13-layers with C1 and C3 terminations, C1C3; and (4) 13-layers with Si1 and Si3 terminations, Si1Si3. The surface energy for these asymmetric slabs can be written as:

$$\gamma_{sum}^{C1Si1} = \gamma_{C1} + \gamma_{Si1} = (E_{slab}^{C1Si1} - n_{Si}^{C1Si1}\mu_{Si}^{SiC} - n_{C}^{C1Si1}\mu_{C}^{SiC})/A$$
$$\gamma_{sum}^{C3Si3} = \gamma_{C3} + \gamma_{Si3} = (E_{slab}^{C3Si3} - n_{Si}^{C3Si3}\mu_{Si}^{SiC} - n_{C}^{C3S3}\mu_{C}^{SiC})/A$$
$$\gamma_{sum}^{C1C3} = \gamma_{C1} + \gamma_{C3} = (E_{slab}^{C1C3} - n_{Si}^{C1C3}\mu_{Si}^{SiC} - n_{C}^{C1C3}\mu_{C}^{SiC})/A$$
$$\gamma_{sum}^{Si1Si3} = \gamma_{Si1} + \gamma_{Si3} = (E_{slab}^{Si1Si3} - n_{Si}^{Si1Si3}\mu_{Si}^{SiC} - n_{C}^{Si1Si3}\mu_{C}^{SiC})/A$$

Here $\gamma_{C1}$, $\gamma_{Si1}$, $\gamma_{C3}$, and $\gamma_{Si3}$ are the surface energies of C1, Si1, C3, and Si3 surfaces; $E_{slab}^{X}$ is the total energy of each slab (X=C1Si1, C3Si3, C1C3, and Si1Si3) with dipole correction, $n_{Si}^{X}$ and $n_{C}^{X}$ are the number of Si and C species within these slabs. Meanwhile, it is known that the surface formation with extra dangling bonds requires more energy, indicating that the surface with three-dangling-bonds per each surface atom is energetically unfavorable, so in the following calculations, we will estimate the energy of surface C1 and Si1 (see in Table S1):

$$\gamma_{C1} = \gamma_{sum}^{C1Si1} \times \frac{\gamma_{sum}^{C1C3}}{\gamma_{sum}^{C1C3} + \gamma_{sum}^{Si1Si3}}$$

$$\gamma_{Si1} = \gamma_{sum}^{C1C3} \times \frac{\gamma_{sum}^{C1Si1}}{\gamma_{sum}^{C1Si1} + \gamma_{sum}^{C3Si3}}$$

*Table S1. The calculated surface energy and surface energy density of SiC surfaces*

| Surface | Surface energy (eV/Å²) | | Surface energy density (eV/atom) | |
|---|---|---|---|---|
| | Si-rich | C-rich | Si-rich | C-rich |
| Si(100) | 0.172 | 0.187 | 1.646 | 1.796 |
| C(100) | 0.236 | 0.221 | 2.264 | 2.114 |
| Step-Si(100) | 0.177 | 0.192 | 1.354 | 1.474 |
| Step-C(100) | 0.322 | 0.306 | 2.466 | 2.346 |
| SiC(110) | 0.190 | 0.190 | 1.286 | 1.286 |
| Si(111) | 0.250 | 0.265 | 2.074 | 2.200 |
| C(111) | 0.452 | 0.433 | 3.755 | 3.596 |

3. **Water chemical potential determination**

The water chemical potential has been estimated in the following equation:

$$\mu(H_2O) = \frac{\langle E(H_2O)_N \rangle}{N} + \Delta E_P - \Delta C_S$$

Here, $\langle E(H_2O) \rangle$ is the ensemble average of the total internal energy of pure water at different temperature, which can be obtained from the AIMD simulations. $N$ is the number of $H_2O$ molecules in the pure water environment, $\Delta E_P$ is an energy correction for pressure deviations in the AIMD from the target pressure appropriate for and approximately incompressible material, $V\Delta P$, and $\Delta C_S$ is the entropy corrosion for water, describing the change in entropy of one $H_2O$ molecule from its standard state to the dissolved phase, which can be obtained from the JANAF database. Note that the target pressure below are set by the pressure of water calculated from the International Association for the Properties of Water and Steam (IAPWS) Formulation 1995 for the Thermodynamic Properties of Ordinary Water Substance for General and Scientific Use [1].

*Table S2. The thermodynamic properties of water calculated from AIMD simulations and from the thermodynamic database.*

| Temperature [K] | Density [g/cm$^3$] | Simulated pressure [kbar] | $\langle E(H_2O)_N \rangle / N$ [eV] | Target pressure [kbar] [1] | $\Delta C_S$ [eV] | $\Delta E_P$ [eV] | $\mu(H_2O)$ [eV] |
|---|---|---|---|---|---|---|---|
| 300 | 1.08 | 5.28 (1.42) | -14.64 (0.02) | ~2.00 | 0.22 | 0.003 | -14.86 (0.02) |
| 400 | 0.96 | 8.91 (0.85) | -14.51 (0.01) | ~0.45 | 0.38 | 0.007 | -14.88 (0.01) |
| 500 | 0.86 | 9.38 (0.57) | -14.38 (0.01) | ~0.45 | 0.56 | 0.007 | -14.93 (0.01) |
| 600 | 0.72 | 8.53 (0.67) | -14.23 (0.01) | ~0.45 | 0.77 | 0.006 | -14.99 (0.01) |
| 700 | 0.48 | 6.18 (0.19) | -14.10 (0.01) | ~0.50 | 1.02 | 0.004 | -15.12 (0.01) |
| 800 | 0.24 | 3.14 (0.15) | -13.95 (0.01) | ~0.50 | 1.33 | 0.002 | -15.28 (0.01) |

[1] H. H. Allan, Thermodynamic Properties of Water: Tabulation from the IAPWS Formulation 1995 or the Thermodynamic Properties of Ordinary Water Substance for General and Scientific Use, https://www.nist.gov/srd/nistir-5078.